\documentclass[a4paper]{article}

\usepackage{INTERSPEECH2020}
\usepackage{cite}
\usepackage{url}
\usepackage{graphicx}
\usepackage{hyperref}
\usepackage{verbatim}
\usepackage{multirow}
\usepackage[table,xcdraw]{xcolor}
\usepackage{array}
\usepackage{tabularx}
\newcolumntype{Y}{>{\centering\arraybackslash}X}

\title{A Cyclical Post-filtering Approach to Mismatch Refinement of Neural Vocoder for Text-to-speech Systems}
\name{Yi-Chiao Wu$^{1}$, Patrick Lumban Tobing$^1$, Kazuki Yasuhara$^1$, Noriyuki Matsunaga$^2$, Yamato Ohtani$^2$, and Tomoki Toda$^{3}$}

\address{
  $^1$Graduate School of Informatics, Nagoya University, Japan\\
  $^2$AI, Inc., Japan\\
  $^3$Information Technology Center, Nagoya University, Japan
 }
\email{yichiao.wu@g.sp.m.is.nagoya-u.ac.jp, ohtani@ai-j.jp, tomoki@icts.nagoya-u.ac.jp}

\begin{document}
\maketitle
\begin{abstract}
Recently, the effectiveness of text-to-speech (TTS) systems combined with neural vocoders to generate high-fidelity speech has been shown. However, collecting the required training data and building these advanced systems from scratch are time and resource consuming. An economical approach is to develop a neural vocoder to enhance the speech generated by existing or low-cost TTS systems. Nonetheless, this approach usually suffers from two issues: 1) temporal mismatches between TTS and natural waveforms and 2) acoustic mismatches between training and testing data. To address these issues, we adopt a cyclic voice conversion (VC) model to generate temporally matched pseudo-VC data for training and acoustically matched enhanced data for testing the neural vocoders. Because of the generality, this framework can be applied to arbitrary TTS systems and neural vocoders. In this paper, we apply the proposed method with a state-of-the-art WaveNet vocoder for two different basic TTS systems, and both objective and subjective experimental results confirm the effectiveness of the proposed framework.
\end{abstract}

\section{Introduction}

Text-to-speech (TTS) is a technique to generate speech according to the given text. Benefitting from the thriving development of neural network (NN), the advanced TTS systems with NN-based waveform generation models~\cite{tacotron2, clarinet} achieve very impressive speech fidelity. However, the high quality and quantity requirements of training data, the burden of data pre-processing, and the time and resource consuming training process make it difficult to build an advanced TTS system from scratch.

A flexible and economical approach to developing a high-quality TTS system is to enhance the speech generated by low-cost or existing TTS systems using an NN-based generation model such as the WaveNet (WN)~\cite{wavenet, sd_wn_vocoder, si_wn_vocoder, ns_wn_vocoder} vocoder. However, there are two challenges for training and testing the NN-based vocoder. First, if the NN-based vocoder is trained with natural acoustic features and waveforms, it will suffer from the acoustic mismatch problem in the testing stage. The acoustic mismatch between the synthetic testing acoustic features, which are extracted from the TTS-generated waveforms, and the natural training acoustic features causes a significant speech quality degradation. Secondly, even if training the vocoder with the synthetic acoustic features and the natural waveforms, the temporal structure mismatch between TTS-generated and natural waveforms still degrades the performance of the vocoder.

To tackle these problems, a cycle-voice conversion (Cycle-VC)~\cite{cyc_vc} model is adopted to respectively generate temporally matched pseudo converted acoustic features for training the NN-based vocoder and acoustically matched enhanced acoustic features in the testing stage. Specifically, the Cycle-VC model includes two conversion paths. The first path converts the synthetic acoustic features to the natural ones, and the second path is composed of a natural to synthetic conversion model following the synthetic to natural conversion model of the first path. The enhanced and the pseudo converted acoustic features can be respectively attained from the first and second paths. Because both the enhanced and the pseudo converted acoustic features are converted by the Cycle-VC model, their acoustic mismatches are less than that of the synthetic and natural acoustic features. Since the pseudo converted acoustic features are converted from the natural acoustic features, their temporal structures are matched to the natural waveforms.

Both objective and subjective evaluations are conducted. The experimental results show the speech quality degradations caused by the acoustic and temporal mismatches and the effectiveness of the proposed framework. To sum up, the contributions of this paper are three folds:
\begin{itemize}
\item This paper argues that TTS-generated speech with manually determined phoneme alignment still has very different temporal structures from the related natural speech, and these temporal mismatches cause significant speech quality degradations.
\item A WN vocoder trained and tested with the proposed framework does enhance the TTS-generated speech.
\item The proposed framework can be generalized for arbitrary TTS systems and neural generation models.
\end{itemize}

\section{Related work}

For TTS systems with an NN-based vocoder, Tacotron2~\cite{tacotron2} has shown an early success by independently training an autoregressive (AR) mel-spectral predictor and then training a WN vocoder with the output of the well-trained mel-spectral predictor. ClariNet~\cite{clarinet} improved it with a non-AR parallel WN-like~\cite{pwn} vocoder and a jointly training manner. The authors of~\cite{gan_tts_wn} also proposed a generative adversarial network~\cite{gan} (GAN)-based framework to jointly optimize its mel-spectral predictor and vocoder. However, these methods are exclusive for specific mel-spectral predictors and difficult to be combined with arbitrary existing TTS systems.

Furthermore, GAN-based~\cite{segan} and WN-based~\cite{wn_se_2017, wn_se_2018} denoising models also have been proposed to directly operate the speech enhancement in the waveform domain. However, because the noisy and clean training data are usually paired, which have matched temporal structures, directly applying these methods for TTS post-filters still has a temporal mismatch problem.

In addition, the learning-based post-filters for synthetic speech enhancement have been explored in different acoustic feature domains and NN-based models~\cite{dnn_pf_2014, dnn_pf_2015, dnn_spss, gan_pf_1, gan_pf_2}. An advanced end-to-end GAN-based postfilter~\cite{wave_cyc_gan_1, wave_cyc_gan_2} also has been proposed to directly generate the enhanced waveforms. Although the GAN-based approaches are effective for addressing the temporal and acoustic mismatches problems, stably training a GAN-based model is still difficult. 

On the other hand, because of the different data length nature of the source and target data in VC, NN-based vocoders trained or fine-tuned with pseudo converted data have been proposed. For instance, the intra-speaker VC frameworks, which can obtain the pseudo converted data for fine-tuning NN-based vocoders, have been explored with Gaussian mixture model~\cite{vc_wn_gmm}, long short-term memory~\cite{vc_wn_lstm}, variational autoencoder~\cite{vc_wn_vae}, and cyclic gated recurrent unit (GRU)~\cite{cyc_vc} models. In this paper, inspired by the VC works combined with the fine-tuned WN vocoders, the pseudo conversion mechanism is applied to the TTS post-filtering scenario.

\begin{figure}[t]
\begin{center}
\includegraphics[width=0.85\columnwidth]{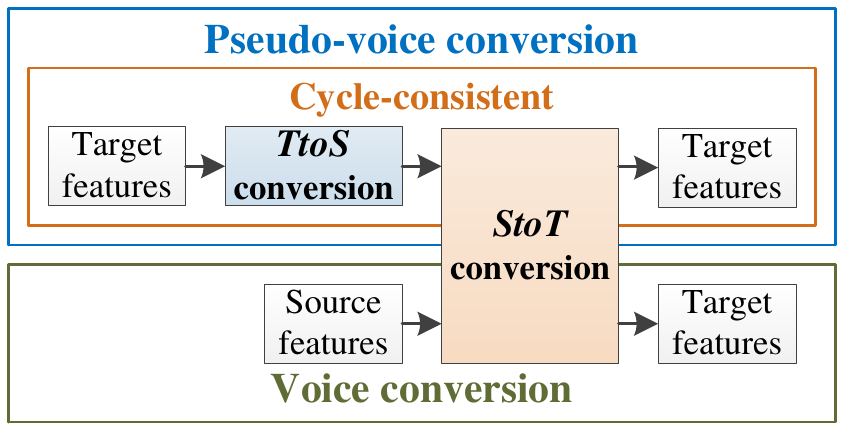}
\caption{Cycle-VC system}
\label{fig:cycvc}
\vspace{2mm}
\includegraphics[width=0.75\columnwidth]{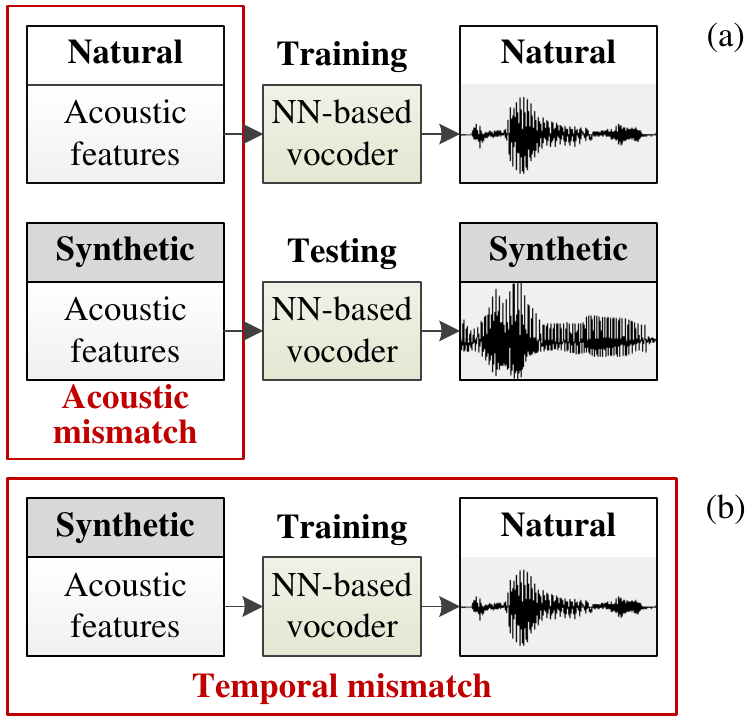}
\caption{Acoustic and temporal mismatches}
\label{fig:am_tm}
\end{center}
\vspace{-10mm}
\end{figure}

\section{Cycle-spectral conversion}

As shown in Fig.~\ref{fig:cycvc}, the Cycle-VC system is composed of a target-to-source ({\it TtoS}) model and a source-to-target ({\it StoT}) model. The conventional VC system usually consists of only a {\it StoT} model, but the Cycle-VC system adopts an additional {\it TtoS} model to advance the speech modeling capability of the {\it StoT} model with the cycle-consistency. Moreover, the self-converted target features are suitable for training or fine-tuning the NN-based vocoders. That is, these self-converted target features are alignment-free to the target waveforms, and their acoustic characteristics are similar to the converted features. 

In this paper, a cycle-spectral conversion model is adopted. Given a source spectral vector $\boldsymbol{X}=\left [ \boldsymbol{x}_{1}^{\top},\cdots, \boldsymbol{x}_{n}^{\top} \right  ]^{\top}$, a target spectral vector $\boldsymbol{Y}=\left [ \boldsymbol{y}_{1}^{\top},\cdots, \boldsymbol{y}_{n}^{\top} \right  ]^{\top}$, an {\it StoT} nonlinear function $f$, and a {\it TtoS} nonlinear function $g$, the loss function is formulated as 
\begin{align}
\mathop{\arg\min}_{\theta, \phi}(\|f(\boldsymbol{X})-\boldsymbol{Y}\|_{L1} + \rho\|f(g(\boldsymbol{Y}))-\boldsymbol{Y}\|_{L1}),
\label{eq:cycvc}
\end{align}
where $\theta$ and $\phi$ are the model parameters of the {\it StoT} and {\it TtoS} models, respectively. $\left \| \cdot  \right \|_{L1}$ is the L1 norm. $\rho$ is a hyper-parameter, which is empirically set to $1e^{-8}$, to avoid the network being dominated by the self-conversion. The network structure consists of input convolution neural network (CNN) layers, AR-GRU blocks, and output CNN layers to convert the spectral features in a framewise manner.

\begin{figure}[t]
\begin{center}
\includegraphics[width=1.0\columnwidth]{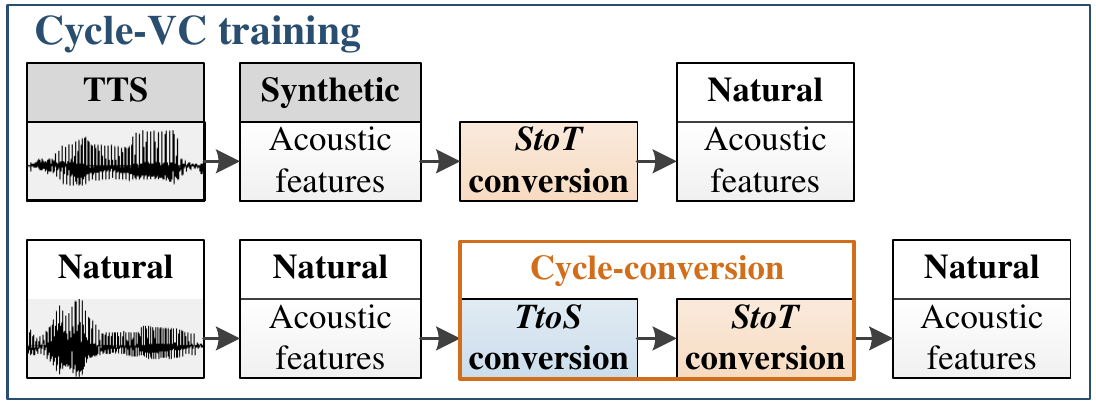}
\caption{Cycle-VC training stage}
\label{fig:cycvc_train}
\end{center}
\vspace{-10mm}
\end{figure}

\section{Proposed post-filter for TTS}

\subsection{Acoustic and temporal mismatches}
Because of the data-driven nature, NN-based vocoders are vulnerable to unseen testing data~\cite{nu_np_2018, cl_2018, cl_2020}. Specifically, NN-based vocoders are usually trained with a pair of natural acoustic features and waveforms, but the input acoustic features in the testing stage are predicted from other models as shown in Fig.~\ref{fig:am_tm} (a). The acoustic mismatch between training and testing data causes significant speech quality degradation.

In this paper, we argue that even if NN-based vocoders are directly trained with a pair of synthetic acoustic features and natural waveforms as Fig.~\ref{fig:am_tm} (b), the temporal mismatch problem still causes severe quality degradation. Specifically, although directly training the vocoder with synthetic acoustic features and natural waveforms can alleviate the acoustic mismatch problem in the testing stage, the temporal mismatch between them still markedly degrade the vocoder. Even if the synthetic acoustic features are extracted from manually tuned TTS-generated speech, which has synchronized phoneme durations, short pauses, and silence segments to the natural target speech, there are still some different temporal structures between TTS-generated and natural waveforms. These temporal mismatches in the vocoder training stage usually cause severe quality degradations such as mispronunciation.   

\begin{figure}[t]
\begin{center}
\includegraphics[width=0.95\columnwidth]{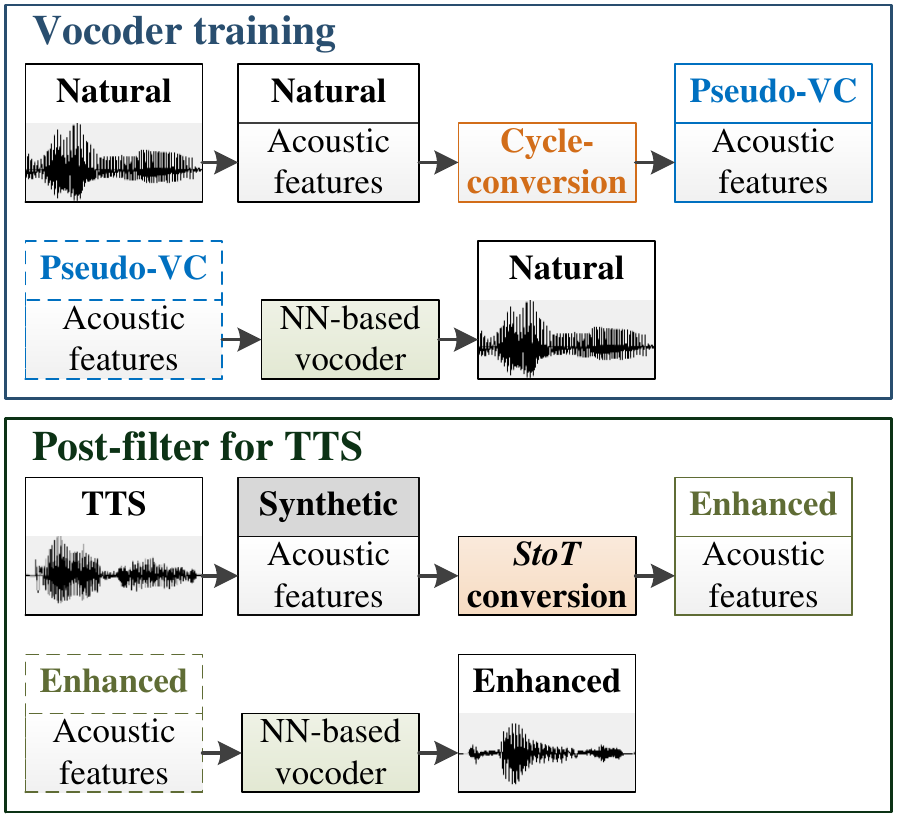}
\caption{Vocoder training and post-filter testing stages}
\label{fig:cycvc_pf}
\end{center}
\vspace{-10mm}
\end{figure}

\subsection{Post-filter with cyclical mismatch refinement}
The proposed method is composed of a Cycle-VC training, a vocoder training, and a post-filter testing stages. As shown in Fig.~\ref{fig:cycvc_train}, synthetic acoustic features, which are extracted from TTS-generated speech, are taken as the source, and natural acoustic features are taken as the target of the Cycle-VC model. The {\it StoT} model is trained with the paired synthetic and natural acoustic features, and the {\it TtoS} model is trained with the cycle-consistency.

As the proposed framework shown in Fig.~\ref{fig:cycvc_pf}, an NN-based vocoder is trained with a pair of temporally matched natural waveforms and pseudo converted acoustic features, which are converted from natural acoustic features using the Cycle-VC model. In the testing stage, the well-trained vocoder generates the enhanced speech from the enhanced acoustic features, which are converted from the synthetic testing acoustic feature using the {\it StoT} model. Note that the pseudo converted and enhanced acoustic features are supposed to be more acoustically matched than the natural and synthetic acoustic features because both of them are converted by the {\it StoT} model.

\section{Experiments}

\subsection{Corpus and TTS system}
An internal Japanese corpus, which included a female and a male speakers, with sampling rate 48~kHz was adopted for developing single-speaker TTS systems. Each speaker had 800 training and 100 testing utterances, and the average length of utterances was around 4 seconds. WORLD-based acoustic features~\cite{world}, which included 60-dimensional mel-cepstral feature ($mcep$), one-dimensional log-scaled fundamental frequency ($F_0$), five-dimensional aperiodicity ($ap$), and their delta and delta-delta terms, were adopted for the TTS systems. The minimum description length was set to 1.0.

Two basic and low-cost TTS systems, Hidden Markov Model (HMM)-based and deep neural network (DNN)-based systems, were adopted, and both of them were trained in a speaker-dependent (SD) fashion using the very limited training data. Specifically, the HMM-based systems were trained with the Hidden Semi-Markov Model (HSMM) training script of HTS (ver. 2.3.2)~\cite{hts}, and the manual phoneme segmentations were adopted to initialize the phoneme HSMMs. The DNN-based systems were composed of four independent feed-forward DNNs for respectively predicting the $F_0$, $ap$, $mcep$, and durations. Both systems adopted a maximum likelihood parameter generation (MLPG)~\cite{vc_mlpg}. The manually refined phoneme segmentations were utilized to generate synthetic speech, which had the same phoneme durations as the natural speech, of the entire training and testing sets. Moreover, a traditional spectral post-filter~\cite{hts, vc_mlpg, hmm_pf}, which was included in the HTS demo, with an enhanced coefficient 1.4 ($\beta=0.4$) was only applied to the DNN-based TTS system. 

\begin{figure}[t]
\begin{center}
\includegraphics[width=0.75\columnwidth]{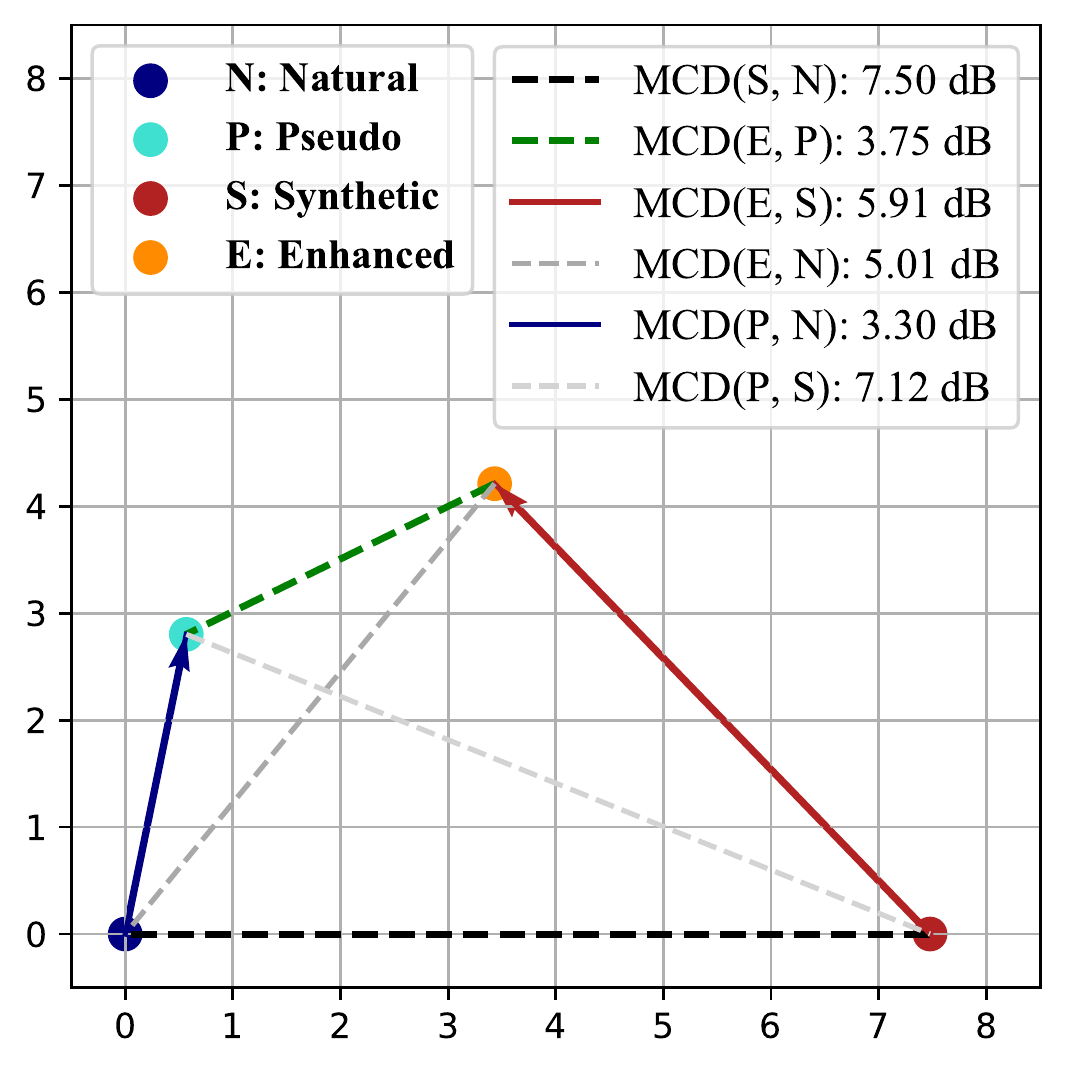}
\caption{Relative mel-cepstral distances of acoustic features on the MCD plane (post-filter w/ DNN-based TTS)}
\label{fig:mcds_dnn}
\vspace{2mm}
\includegraphics[width=0.75\columnwidth]{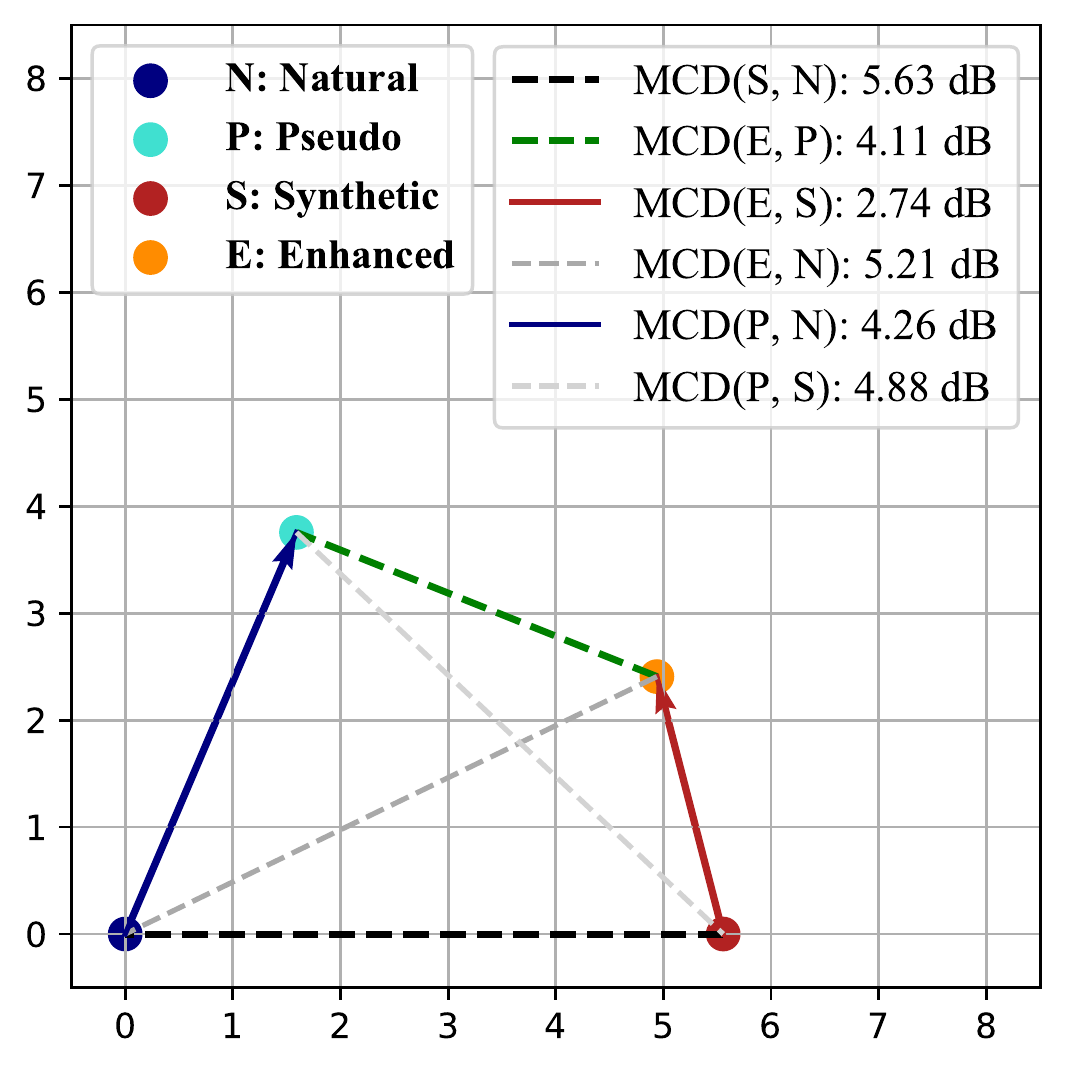}
\caption{Relative mel-cepstral distances of acoustic features on the MCD plane (post-filter w/ HMM-based TTS)}
\label{fig:mcds_hmm}
\end{center}
\vspace{-10mm}
\end{figure}

\subsection{Cycle-VC model and WN vocoder}
Both natural and TTS utterances were downsampled to 24~kHz for SD Cycle-VC models and SD WN vocoders. WORLD-based acoustic features, which included 45-dimensional $mcep$, one-dimensional log-scaled $F_0$ and unvoiced/voiced ($U/V$) binary code, and three-dimensional coded $ap$, were adopted for the input of the Cycle-VC models and the auxiliary features for the WN vocoders. Note that the outputs of the Cycle-VC models were only the $mcep$ features. The settings of the Cycle-VC model followed the previous work~\cite{cyc_vc}, and the training epoch was set to 15. Furthermore, several SD WN vocoders were involved in the evaluations to show the effectiveness of the proposed framework for different speakers and systems. The architecture and training processing of the WN vocoders followed our previous work~\cite{qpnet_2020} with 200,000 iterations. 

\subsection{Objective evaluations}
Figures~\ref{fig:mcds_dnn} shows the relationships among natural (N), synthetic (S), pseudo converted (P), and enhanced (E) $mceps$ of the proposed neural post-filters with DNN-based TTS systems on a mel-cepstral distortion (MCD) plane. The distance between any two points represents the average MCD of them. Specifically, although the TTS and natural speech waveforms have the same contexts and durations, the distance between the natural and synthetic $mceps$ (MCD(S,~N)) is still the longest distance on the plane, and this result implies the temporal mismatches of the TTS and natural speech waveforms and the severe mismatches of their $mceps$. The smaller value of MCD(E,~P) than the value of MCD(S,~N) indicates that the proposed framework does alleviate the acoustic mismatches between the training and testing $mceps$ of NN-based vocoders. The smaller value of MCD(E,~N) than the value of MCD(S,~N) also shows the effectiveness of the Cycle-VC model to enhance the synthetic mcep. Furthermore, Fig.~\ref{fig:mcds_hmm} has a similar tendency as Fig.~\ref{fig:mcds_dnn}, which shows the generality of the proposed framework even with different TTS systems. 

\begin{table}[t]
\caption{Comparison of testing WN vocoders}
\vspace{-2mm}
\label{tb:con}
\fontsize{9pt}{10.8pt}
\selectfont
{%
\begin{tabularx}{\columnwidth}{@{}p{1.5cm}YY@{}}
\toprule
        & \multicolumn{2}{c}{Acoustic features} \\ 
        & Training           & Testing \\ \midrule
Natural &                    & Natural   \\
DNN-AM  & Natural            & Synthetic \\
HMM-AM  &                    & Synthetic \\ \midrule
DNN-TM  & Synthetic          & Synthetic \\ 
HMM-TM  & Synthetic          & Synthetic \\ \midrule
DNN-NPF & Pseudo   converted & Enhanced  \\ 
HMM-NPF & Pseudo   converted & Enhanced  \\ \bottomrule
\end{tabularx}%
}
\end{table}

\subsection{Subjective evaluations}
As shown in Table~\ref{tb:con}, seven training and testing combinations of WN vocoders were included in the subjective evaluations. Specifically, the WN vocoder trained with natural acoustic features was tested by natural and DNN/HMM-based synthetic acoustic features, which were the natural and acoustic mismatch (AM) scenarios, respectively. The WN vocoders trained and tested with the DNN/HMM-based synthetic acoustic features were the temporal mismatch (TM) scenarios. The WN vocoders respectively trained and tested with DNN/HMM-based pseudo converted and enhanced acoustic features were the neural post-filter (NPF) scenarios. The subjective evaluations included the systems with these seven scenarios and the DNN-based and HMM-based TTS systems. Note that all systems were trained in an SD manner, so the total number of systems in the subjective tests was 18.

For each testing system and scenario, we randomly selected 50 testing utterances to form the subjective set and the total number of the utterances was 900. Ten subjects were involved in a preference test and a mean opinion score (MOS) test, and most of them were native speakers. Each subject evaluated a part of the subjective set, and each utterance in the subjective set was at least evaluated by one subject. The final results were the average scores of the testing speakers.  

\begin{figure}[t]
\begin{center}
\includegraphics[width=0.9\columnwidth]{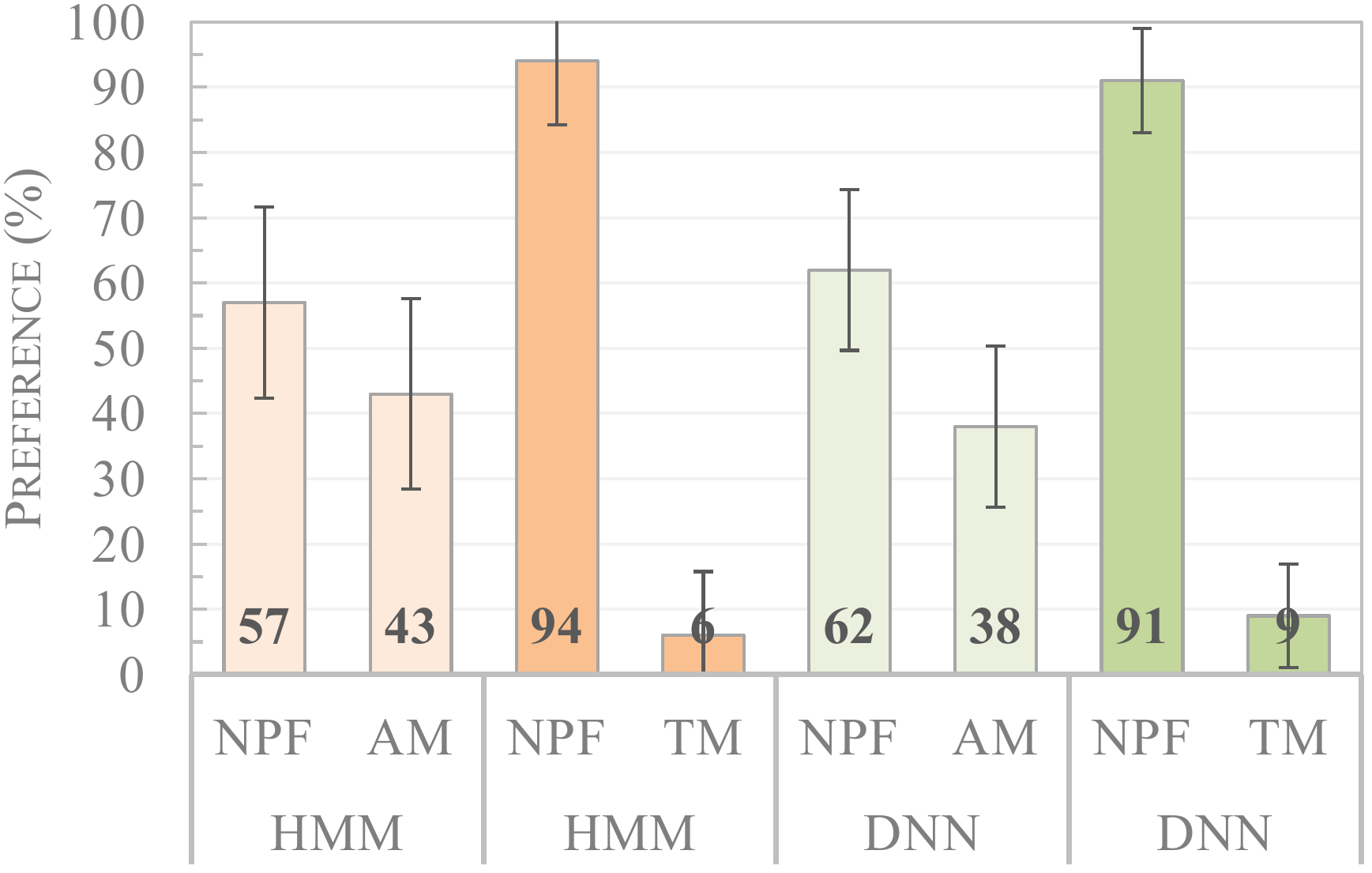}
\caption{Preference results with 95\% confidence interval}
\label{fig:pk}
\vspace{2mm}
\includegraphics[width=0.9\columnwidth]{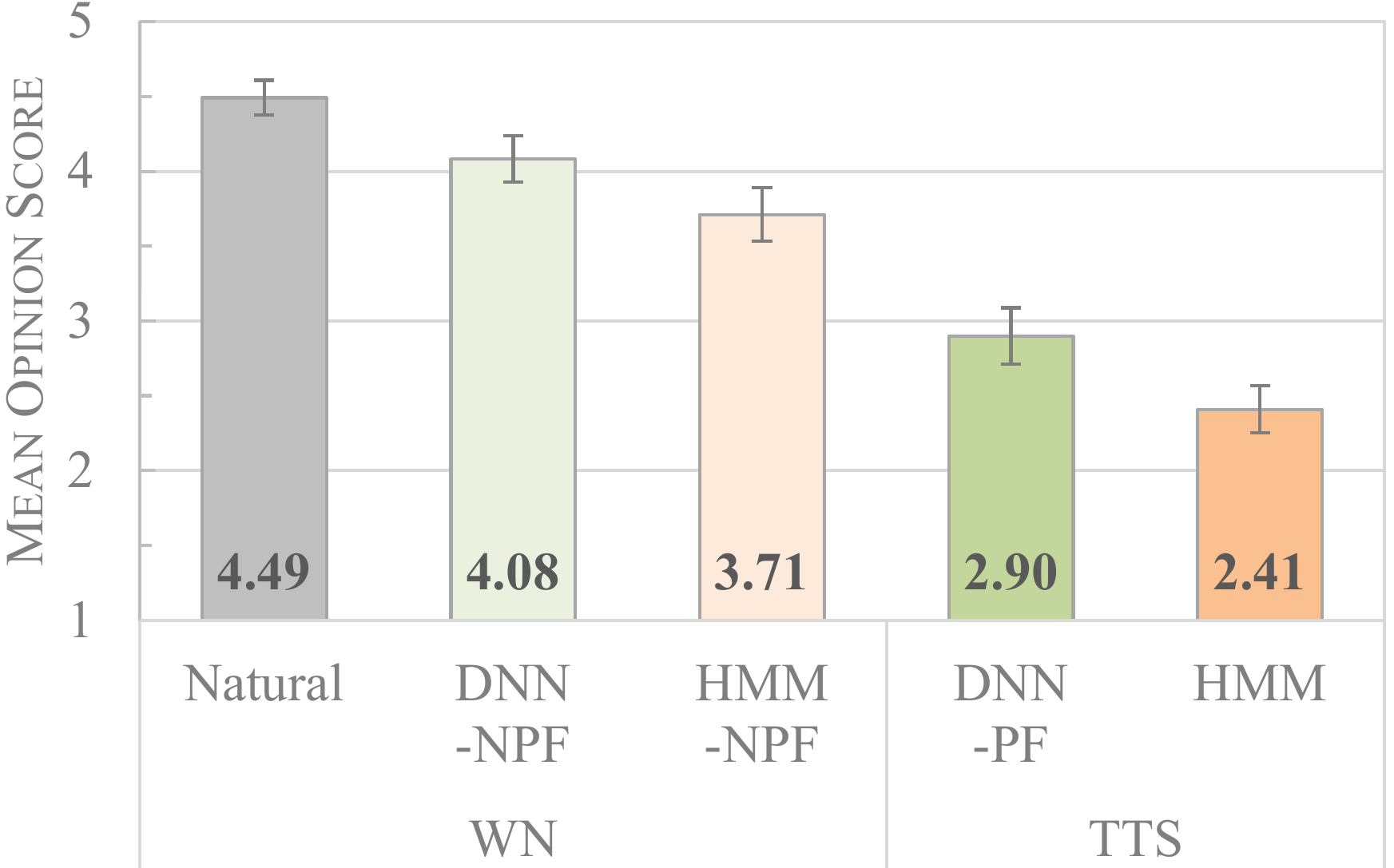}
\caption{MOS results with 95\% confidence interval}
\label{fig:mos}
\end{center}
\vspace{-10mm}
\end{figure}

As shown in Fig.~\ref{fig:pk}, the enhanced utterances, which were generated by the WN vocoders with the proposed post-filtering framework, were respectively compared with the utterances suffering the AM and TM problems. The results show the effectiveness of the proposed framework to alleviate the AM and TM problems and imply the generality of it for different TTS systems. The results also confirm our assumption that although the TTS-generated speech has the same phoneme durations as the natural speech, the different temporal structures still cause a severe TM problem. Moreover, the AM problem also causes significant speech quality degradations according to the results, so the proposed framework is essential for the post-filter application with an NN-based vocoder. 

Figure~\ref{fig:mos} shows the results of the MOS test, where each subject was asked to give a score (1--5) to evaluate the speech quality of the given utterance. The higher the socre, the higher the speech quality. The results show that the proposed post-filter markedly enhanced the TTS-generated speech, even if a traditional spectral post-filter was already applied to the DNN-based TTS. Although there is still a room to improve the speech quality to attain the same quality as the upper bound, which is the WN vocoder trained and tested with the natural acoustic features, the significant improvements ($>1$) of MOSs for both TTS systems still show the effectiveness of the proposed framework as a post-filter for arbitrary TTS systems. 

\section{Conclusions}

In this paper, the harmful effects of the acoustic and temporal mismatches for the TTS post-filter with an NN-based vocoder are explored. The proposed framework adopts the Cycle-VC framework to get the temporally matched pseudo converted acoustic features for the training of the NN-based vocoder and the acoustically matched enhanced acoustic features for the testing of the neural post-filter. Both objective and subjective tests of different TTS systems and speakers show the generality and effectiveness of the proposed framework. For future works, we intend to explore the proposed framework with more different NN-based vocoders.

\section{Acknowledgments}

This work was partly supported by JSPS KAKENHI Grant Number 17H06101 and
JST, CREST Grant Number JPMJCR19A3.

\bibliography{mybib}
\bibliographystyle{IEEEtran}

\end{document}